\documentclass[twocolumn,aps,prl,showpacs,superscriptaddress]{revtex4}
\usepackage{graphics,bm}
\usepackage{amssymb}
\usepackage{epsfig}
\usepackage{epsf}

\def\be{\begin{equation}} \def\ee{\end{equation}}
\def\beq{\begin{eqnarray}} \def\eeq{\end{eqnarray}}

\def\nn{\nonumber}

\begin{document}

\title{Crossover From Strong to Weak Pairing States in $t-J-U$ Model Studied by A Slave Spin Method}
\author{Wei-Cheng Lee}
\email{wlee@binghamton.edu}
\affiliation{Department of Physics, Applied Physics, and Astronomy, Binghamton University - State University of New York, Binghamton, 13902, USA}

\begin{abstract}
We investigate the superconductivity in the $t-J-U$ model within a slave-spin formalism. We show that the BCS mean-field theory implemented with the slave spin formalism naturally predicts 
two distinct gaps which are the pairing gap of the spinons $\Delta_f$ and the Cooper pairing gap of the electrons $\Delta_{SC} = Z\Delta_f$, where $Z$ is the quasiparticle weight.
Furthermore, we find that the nature of the superconducting state depends crucially on the interaction providing the pairing mechanism. For spin interactions, 
the bandwidth of the spinon hopping term is renormalized by $Z$ but the its pairing term is not. 
As a result, if $U$ exceeds the critical value for the Mott insulating state at half-filling, $Z$ develops a strong doping dependence, leading to a doping-driven crossover from strong to weak pairing 
states. In the strong pairing state, while $\Delta_f$ is enhanced as $x\to 0$, $\Delta_{SC}\sim x$ due to the renormalization of $Z$. 
In the weak pairing state, $Z$ does not change with $x$ significantly. Therefore, $\Delta_{SC}$ is mainly controlled by $\Delta_f$, and both of them go to zero at larger doping. 
The crossover from strong to weak pairing states is well captured by the slave spin formalism within reasonable range of parameters just at the mean-field level, indicating the slave
spin formalism is a powerful tool to study correlated materials.
For charge interactions, we find that the bandwidth of the spinon hopping term and its pairing term are renormalized by $Z$ and $Z^2$ respectively. 
Consequently, both $\Delta_f$ and $\Delta_{SC}$ are suppressed at small $x$ in large U, and no crossover will occur. 
The implication of our results for the superconducting states in correlated materials will be discussed.
\end{abstract}


\maketitle

{\it Introduction} --
The physics of the high temperature cuprate superconductors is usually attributed to doping a Mott insulator\cite{leermp2006}. Theoretically, it is believed that the Hubbard model describing the 
strong onsite Coulomb repulsive interaction holds a crucial key to uncloak the mystery of the cuprates.
While the Hubbard model has been shown to capture various experimental results qualitatively\cite{leermp2006}, how the system evolves from strongly to weakly coupled by the doping in the Hubbard 
model remains a challenging question.\cite{phillips2010} 
To account for the $d$-wave superconductivity observed in cuprates, the $t-J$ model derived from a second order perturbation theory on the Hubbard model via the superexchange mechanism 
is often employed to study the relation between antiferromagnetism near zero doping and the $d$-wave superconductivity at finite dopings.\cite{leetk2003,leewc2003,lugas2006,tan2008,ma2013} 
While the $t-J$ model predicts both the antiferromagnetism and the $d$-wave superconducting state naturally at the mean-field level, 
the no-double occupation constraint, which is a requirement of two electrons prohibited on the same site,
is very difficult to handle. Moreover, since this constraint is exact only at infinite $U$, the $t-J$ model usually fails to reproduce the weakly interacting 
limit at larger doping.
In order to overcome these difficulties, the $t-J-U$ model\cite{daul2000,basu2001,zhangfc2003,yuanf2005,abram2013,farrell2014} 
is proposed to be another promising candidate to describe the cuprates, 
in which the notorious no-double occupation constraint is relaxed by considering the onsite Hubbard interaction $U$ directly.
It is then expected that a $t-J-U$ model with a finite but large $U$ could possibly combine all the merits of the Hubbard and the $t-J$ models, but to do so, a reliable treatment 
for the Hubbard $U$ term is necessary, which unfortunately routes back to the original question to be solved.

Recently, the slave spin formalism has been widely used to study the Mott transition in multiorbital systems.
\cite{demedici2005,demedici2009,hassan2010,demedici2011,yu2011,yu2012,yu2013,demedici2014,giovannetti2015,mukherjee2016}
In this formalism, while the physical spin of the electron is described by the same fermionic spinon $f$ used in the slave-boson approach,\cite{kotliar1986,li1989,ubbens1992,leermp2006} 
the charge degrees of freedom are represented by a quantum spin 1/2. 
As a result, the charge fluctuations can be described by spin flips between zero and one charge states, subject to a constraint on each site to remove the 
enlarged Hilbert space.
The advantage of this method is that the dimension of the Hilbert space enlarged by the slave spin is finite, which leads to a more controllable mean-field theory.
Surprisingly, it has been shown that even at the mean-field level, the slave spin formalism can obtain the Mott insulating state in a good agreement with the dynamical mean-field theory (DMFT),
and the quasiparticle weight $Z$ obtained in the large $U$ limit reproduces the famous Gutzwiller approximation $g_t=2x/(1+x)$.\cite{gutzwiller1963,florian1990,hassan2010}

Motivated by these advantanges, we develop a BCS theory implemented with the slave spin formalism to study the superconducting state in the $t-J-U$ model. 
Two distinct gaps, the spinon pairing gap $\Delta_f$ and the superconducting gap $\Delta_{SC} = Z\Delta_f$, arise naturally.
We find that a crossover from strong to weak pairing states driven by the doping $x$ could occur as $U$ exceeds the
critical value $U_c$ for the Mott insulating state at half-filling. 
Moreover, the doping dependences of $\Delta_f$ and $\Delta_{SC}$ are found to differ fundamentally in the strong and the weak pairing states, and a superconducting dome appears as a 
direct consequence of the crossover. 
On the other hand, if the pairing is induced by charge interactions, e.g., nearest-neighbor attractive Coulomb interaction, no crossover would occur.
All the results are obtained at the mean-field level within reasonable ranges of the model parameters, which demonstrates 
that the slave spin formalism is a powerful technique to study physics related to transition from strong to weak coupling states within the same framework.

{\it Model and Formalism} -- We consider a generic single band $t-J-U$ model $H=H_t + H_J + H_U$, where
\beq
H_t&=& \sum_{ij,\sigma} t_{ij} c^\dagger_{i\sigma} c_{j\sigma}\,\,\,,\,\,\,H_U = U\sum_i n_{i\uparrow} n_{i\downarrow},\nn\\
H_J&=& J\sum_{\langle i,j\rangle} \vec{S}_i\cdot\vec{S}_j.
\eeq
$J/t=0.4$ is used for every calculation.
We employ the $U(1)$ version of the slave spin formalism\cite{yu2012} to treat the Hubbard $U$ term, thus the electron creation operator is represented by
\be
c^\dagger_{i\sigma} = S^+_{i\sigma} f^\dagger_{i\sigma},
\ee 
where $f^\dagger_{i\sigma}$ creates a physical spin $\sigma$ at site $i$ and $S^+_{i\sigma}$ creates a charge $-e$ at site $i$. The constraint to project out the unphysical state 
is $S^z_{i\sigma} = f^\dagger_{i\sigma} f_{i\sigma} - 1/2$.
$H_J$ term is decomposed into the $d$-wave superconducting channel by the same Hubbard-Stratonoch fields proposed by Ubbens and Lee.\cite{ubbens1992} 
Because $H_J$ describes interactions between physical spins,
the terms generated from $H_J$ will not involve slave spin degrees of freedom at the mean-field level.
The resulting mean-field pairing term is
\be
H^{MF}_J = -\frac{J'}{2}\sum_{\langle i,j\rangle} \Delta_f^{ij\,*} f_{i\uparrow} f_{j\downarrow} - f_{i\downarrow} f_{j\uparrow} + h.c.,
\ee
where $J' = 3J/4$, and $\Delta_f^{ij}\equiv \langle f_{i\uparrow} f_{j\downarrow} - f_{i\downarrow} f_{j\uparrow}\rangle$
is the spinon pairing order parameter which has the $d$-wave symmetry of $\Delta_{f} = \Delta_f^{i,i+\hat{x}} = - \Delta_f^{i,i+\hat{y}}$.
Next we follow the $U(1)$ slave spin formalism to treat $H_t+H_U$ terms\cite{yu2012}, and the procedure is described briefly below.
We start by performing the standard Hubbard-Stratonoch transformation to decouple the spinon and the slave spin degrees of freedom so that we can write down their Hamiltonians separately. 
Next, we assume that the dynamics of the slave spins is the same for each site, thus the mean-field slave-spin Hamiltonian is reduced to a single site problem which can be diagonalized exactly.
Finally, we solve the corresponding mean-field equations subject to the constraint which is satisfied on the average at the mean-field level.
 
The final mean-field Hamiltonians are
\beq 
H^{f,MF}&=& \sum_{\vec{k}}\big[\sum_{\sigma} \big(Z_\sigma \epsilon(\vec{k}) -\mu\big) f^\dagger_{\vec{k}\sigma}
f_{\vec{k}\sigma} \big]\nn\\
&-& \big[\Delta(\vec{k}) f_{\vec{k}\uparrow} f_{-\vec{k}\downarrow} + h.c.\big],\nn\\
H^{s,MF} &=& \sum_\sigma \big(\epsilon^{\sigma\sigma} \langle \tilde{z}_\sigma\rangle \tilde{z}^\dagger_\sigma + h.c.\big) + \lambda_\sigma S^z_\sigma\nn\\
&+& U\big(S^z_\uparrow + \frac{1}{2}\big)\big(S^z_\downarrow + \frac{1}{2}\big),
\label{mf}
\eeq
where $Z_\sigma\equiv \vert \langle \tilde{z}_\sigma\rangle\vert^2$ is the quasiparticle weight, $\epsilon(\vec{k}) = 2t(\cos k_x + \cos k_y)$, 
$\Delta(\vec{k}) = J'\Delta^*_f(\cos k_x - \cos k_y)$, and $\tilde{z}_\sigma = S^-_\sigma/\sqrt{(1/2+\delta)^2 - \vert\langle S^z_\sigma\rangle\vert^2}$ where $\delta$ is an infinitesimal value to 
regularize $\langle \tilde{z}_\sigma\rangle$ for the case of $\langle S^z_\sigma\rangle = 1/2$. 
$\epsilon^{\sigma\sigma}=\frac{1}{\Omega}\sum_{\vec{k}} \epsilon(\vec{k}) \langle f^\dagger_{\vec{k},\sigma} f_{\vec{k},\sigma}\rangle$ is the average kinetic energy of the spinon with spin $\sigma$.
The parameters of $\Delta_f$, $Z_\sigma$, $\mu$, and $\lambda_\sigma$ are obtained by solving the corresponding mean-field equations subject to the following constraint
\be
1-x = \frac{1}{\Omega}\sum_{\vec{k},\sigma}\langle f^\dagger_{\vec{k},\sigma} f_{\vec{k},\sigma}\rangle = 1 + \sum_\sigma\langle S^z_\sigma\rangle,
\ee
where $x$ is the doping away from the half-filling.
The derivation of the mean-field equations can be found in Supplementary Materials. In the present work, we only focus on spin singlet pairing state, and consequently all the parameters 
are spin-independent. We will drop the $\sigma$ index from now on.

\begin{figure}
\includegraphics[width=2.6in]{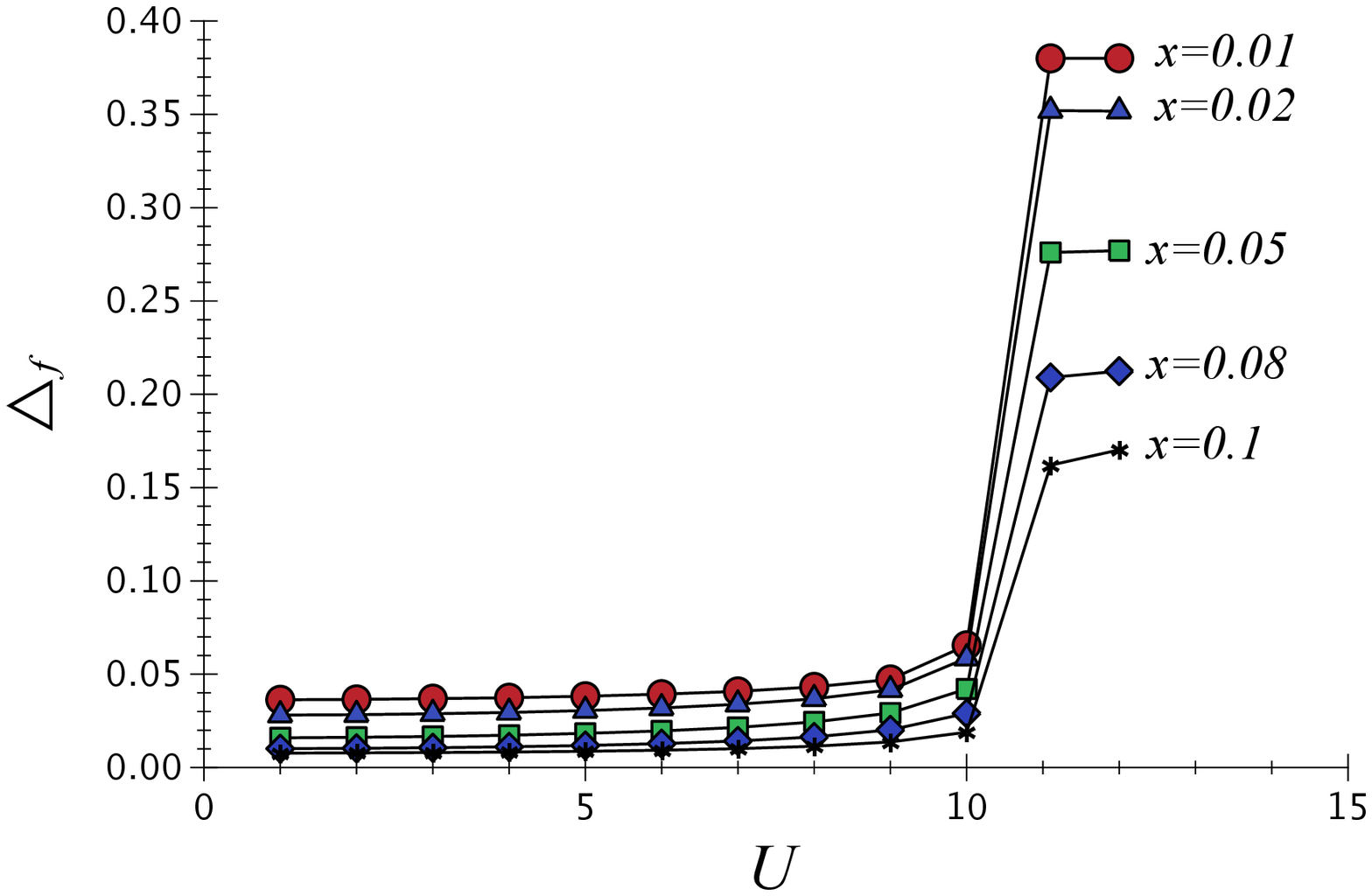}
\includegraphics[width=2.6in]{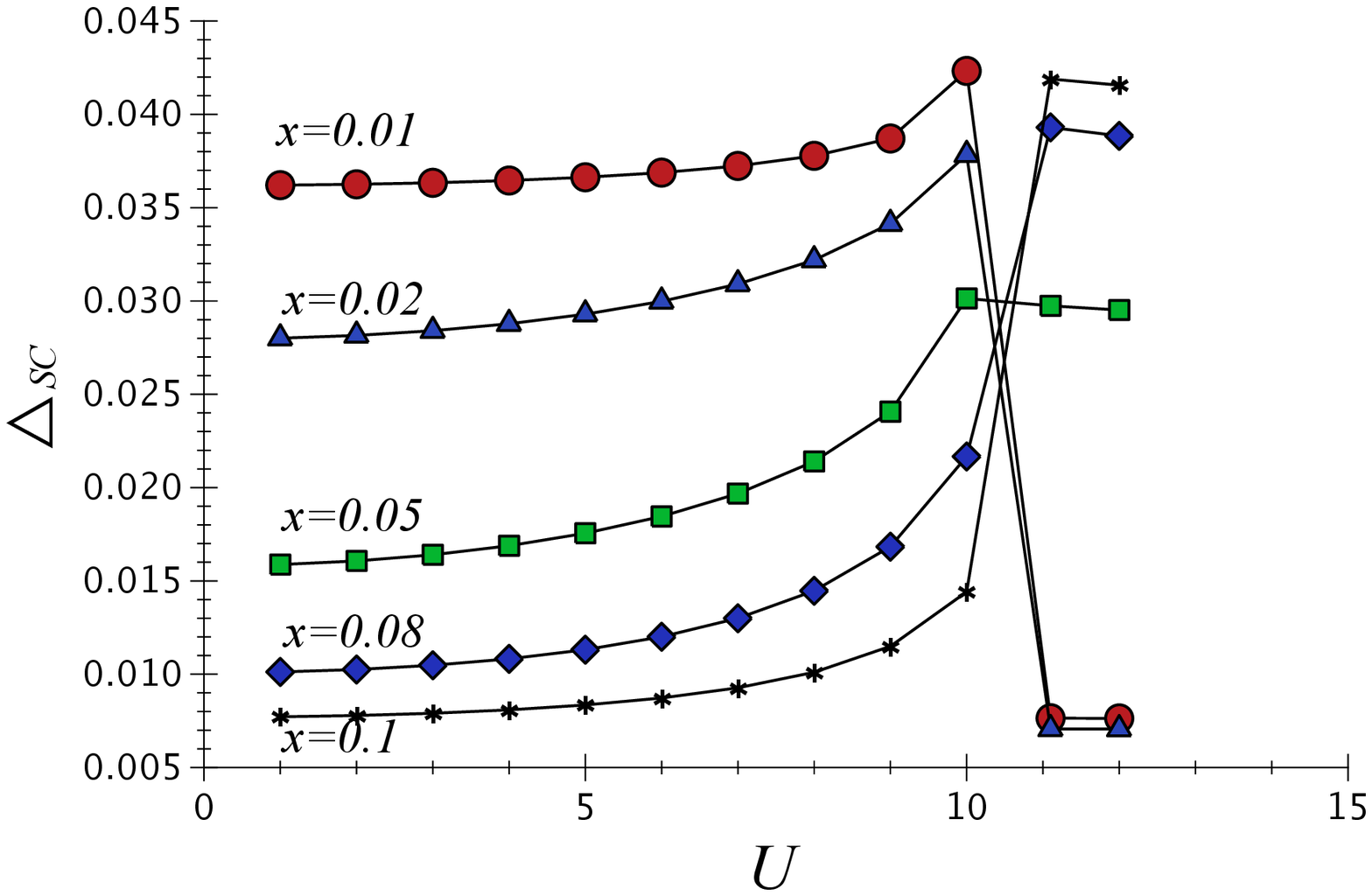}
\includegraphics[width=2.6in]{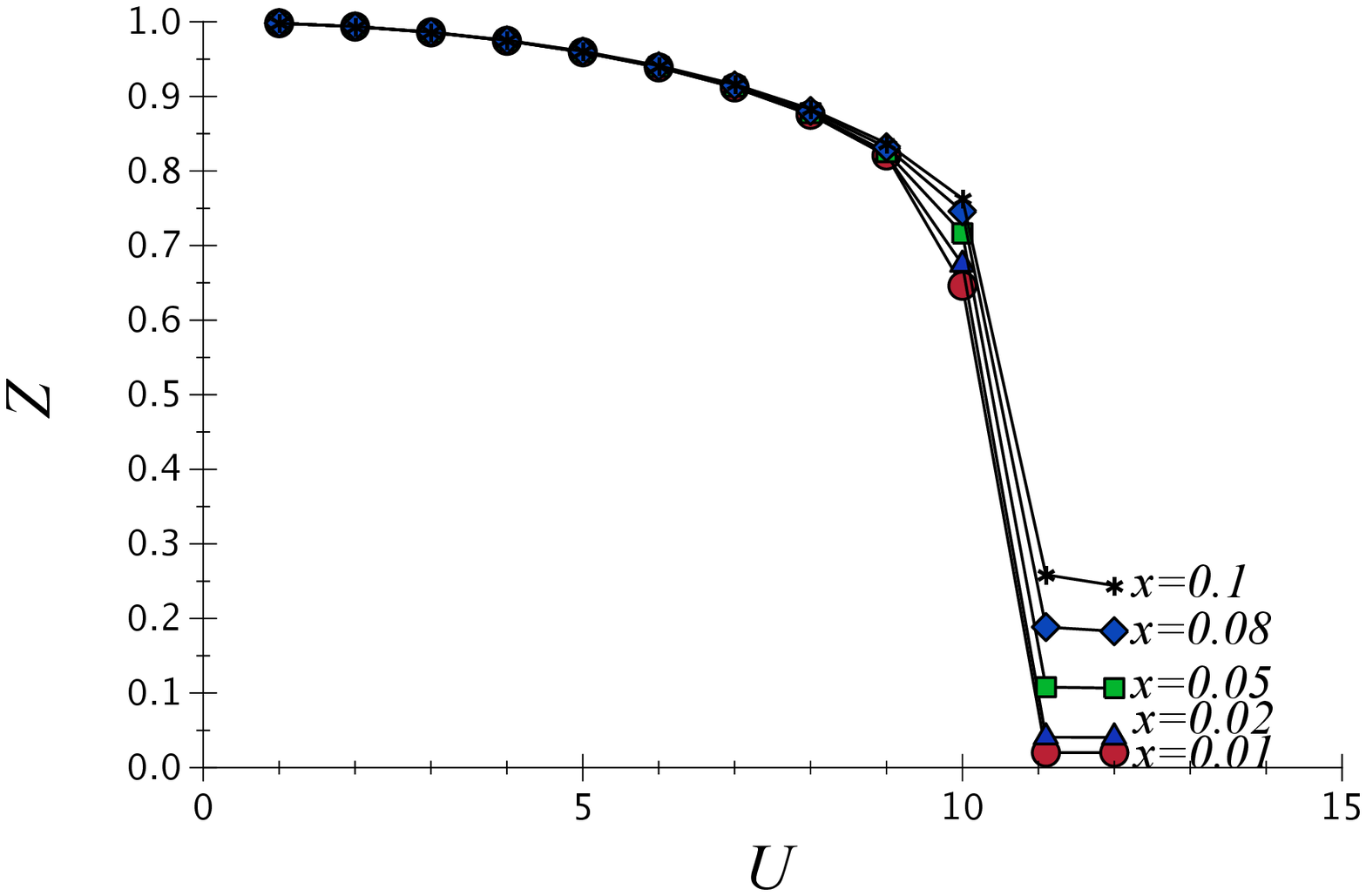}
\caption{\label{fig:one} (a) The spinon pairing order parameter $\Delta_f$ as a function of $U$ for different doping $x$. $\Delta_f$ increases with the increase of $U$ and it is always 
larger at small $x$. An abrupt increase of $\Delta_f$ occurs as $U/t>11$ due to the abrupt decrease of $Z$. indicating a crossover to the strong pairing state. 
(b) The electron pairing order parameter $\Delta_{SC}$ as a function of $U$ for different doping $x$. In the strong pairing state ($U/t > 11$), $\Delta_{SC}$ is limited by $Z$, thus $\Delta_{SC}\sim x$ 
at small $x$. (c) The quasiparticle weight $Z$ as a function of $U$ for different doping $x$.}
\end{figure}

\begin{figure}
\includegraphics[width=2.6in]{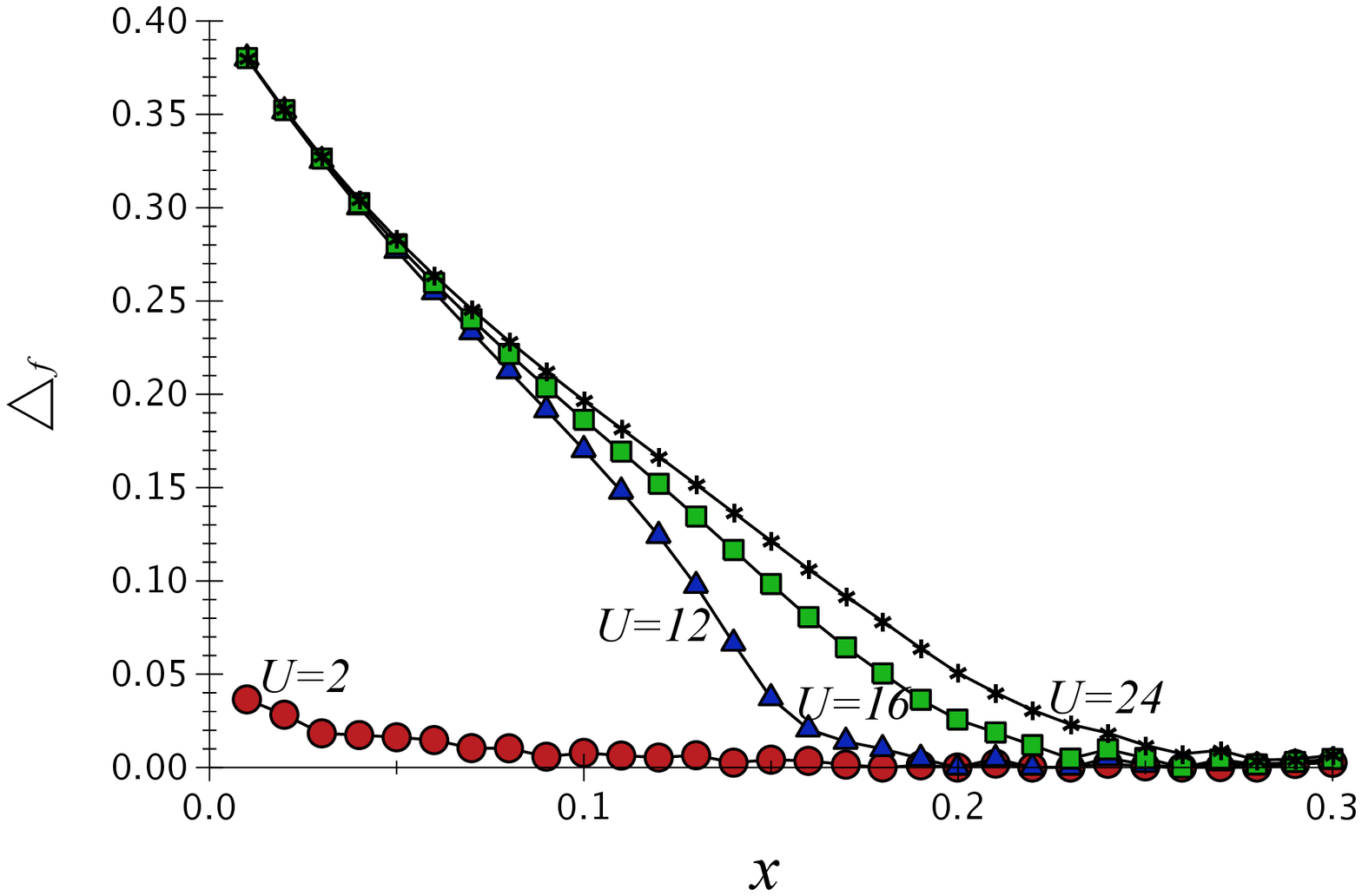}
\includegraphics[width=2.6in]{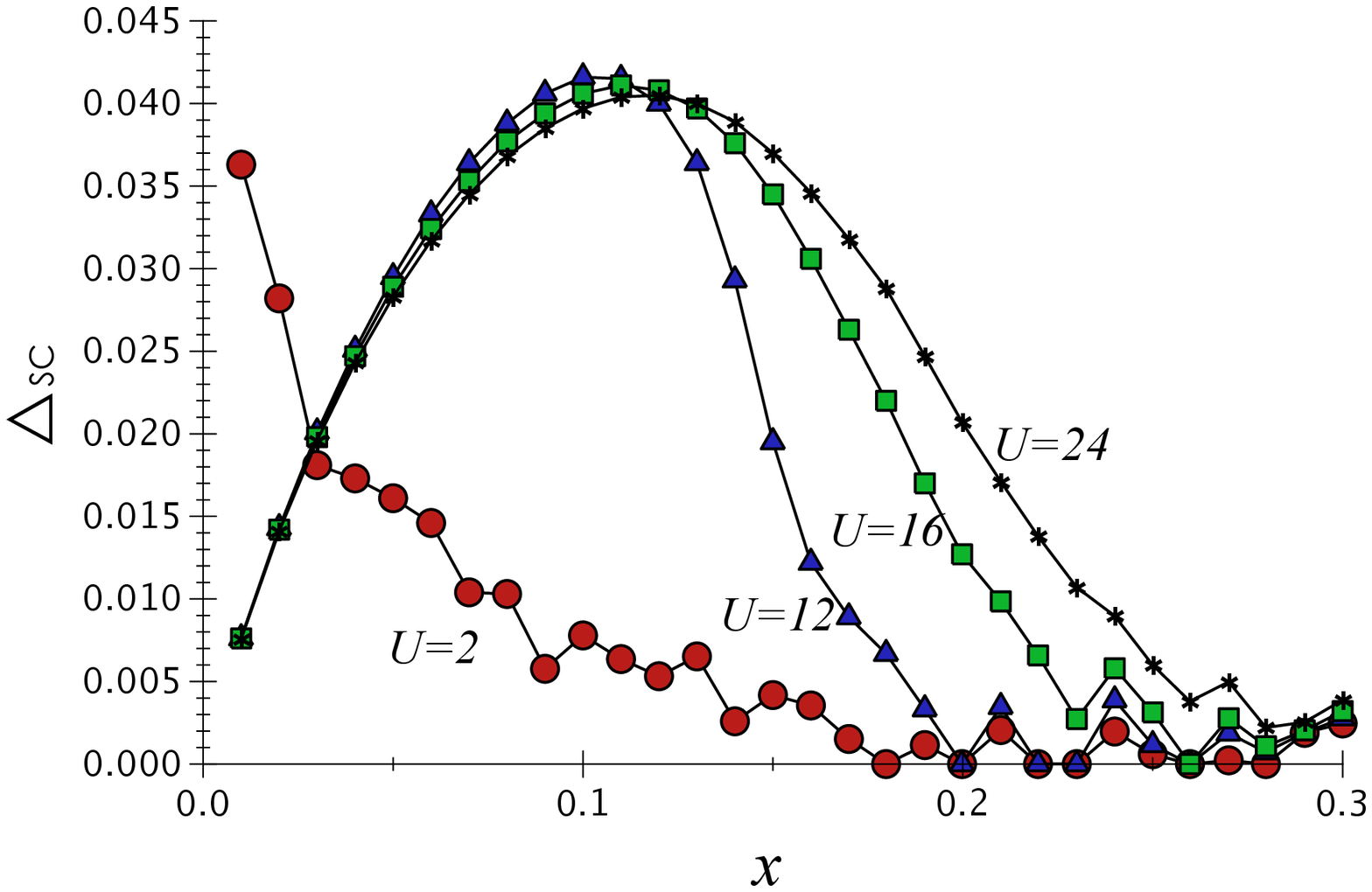}
\includegraphics[width=2.6in]{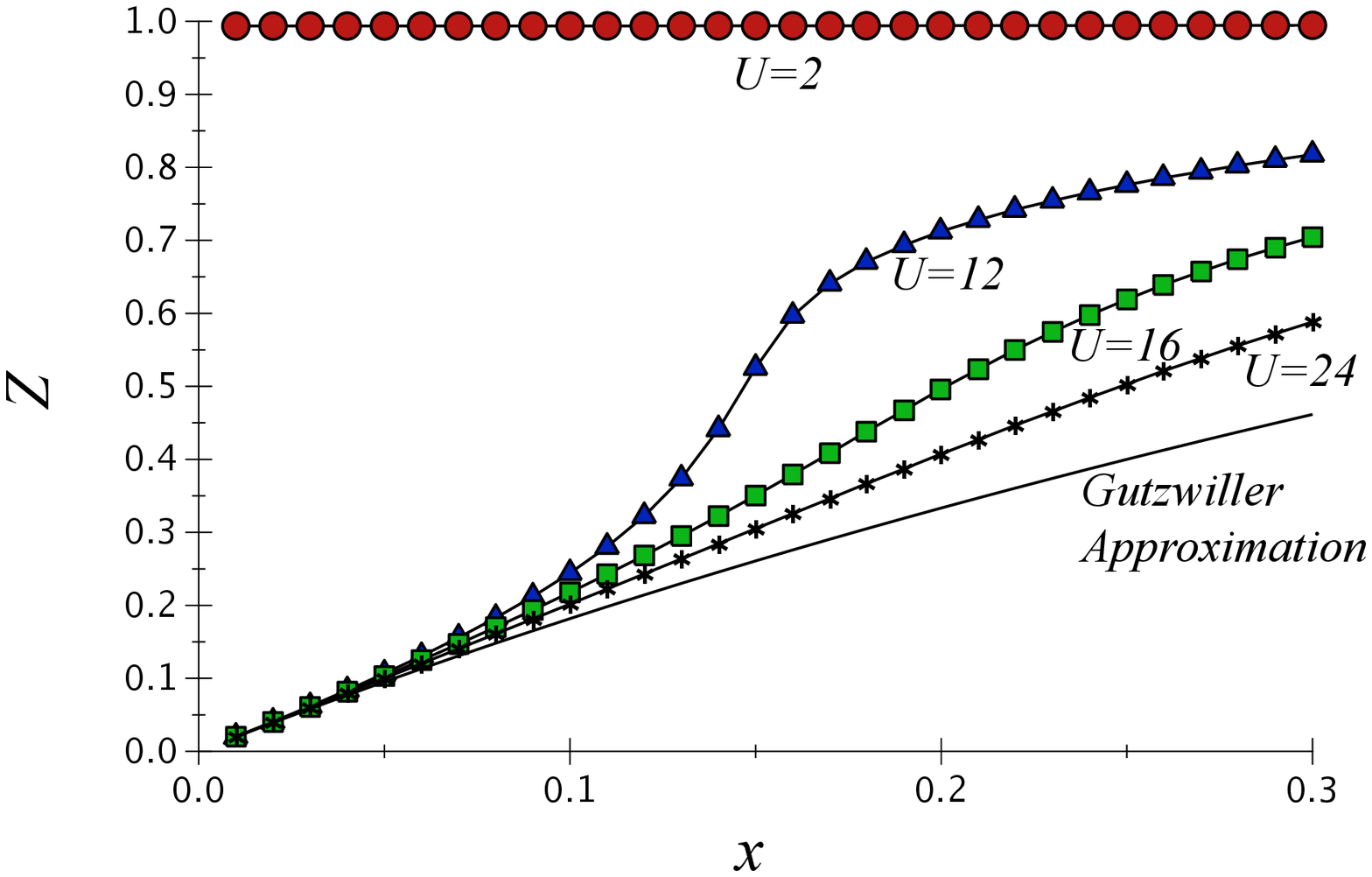}
\caption{\label{fig:two} (a) The spinon pairing gap $\Delta_f$ as a function of $x$ for $U/t=2,12,16,24$. The case of $U=2$ represents the weak pairing limit. 
Other cases exhibit a strong enhancement with the decrease of $x$.
(b) The corresponding electron pairing gap $\Delta_{SC}$ as a function of $x$. In the weak pairing limit ($U=2$), $\Delta_{SC}$ decreases 
with $x$. Other cases exhibit the crossover from strong to weak pairing states with the decrease of $x$. At small $x$, $\Delta_{SC}$ is limited by $Z$ so that $\Delta_{SC}\sim x$. At larger
$x$, $\Delta_{SC}$ is mainly determined by the spinon pairing $\Delta_f$, thus $\Delta_{SC}$ starts to decrease after a critical doping. It is interesting to note that a superconducting dome
appears naturally. (b) The corresponding quasiparticle weight $Z$ as a function of $x$. The black line represents the Gutzwiller factor $g_t=2x/(1+x)$. The slave spin matches $g_t$ at small
doping very well.}
\end{figure}

Once the mean-field equations are solved, we can obtain the electron pairing order parameter, which is also the true superconducting order parameter, by
\be
\Delta_{SC} = \langle S^-_{i\uparrow}S^-_{j\downarrow}\rangle \Delta_f\approx Z \Delta_f
\label{dsc}
\ee
{\it Results} --
It is instructive to discuss the limits of $U\to 0$ and $U\to\infty$. For $U\to 0$, $Z\approx 1$ and $\Delta_f\approx \Delta_{SC}$. As a result, we recover the standard mean-field results on 
$t-J$ model without no-double occupation constraint, 
which only has single gap and the largest gap is found at the half-filling. On the other hand, for $U\to \infty$, $Z$ reduces to the Gutzwiller factor $g_t=2x/(1+x)$ and 
$\Delta_f$ and $\Delta_{SC}$ become two distinct gaps. From $H^{f,MF}$ in Eq. \ref{mf}, because the kinetic energy term is renormalized by $Z$ while the pairing term is not, the spinon pairing 
$\Delta_f$ is greatly enhanced at small $x$, and the effect of $Z$ becomes insignificant at larger $x$. 
In contrast, because the electron pairing gap $\Delta_{SC}$ is given by Eq. \ref{dsc}, $\Delta_{SC}\sim x$ at small $x$ and $\sim \Delta_f$ at large $x$. 
As a result, a crossover from strong to weak pairing states driven by the doping $x$ is expected in the large $U$ limit, which leads to a superconducting dome naturally.
The mean-field results demonstrating the crossover are plotted in Fig. \ref{fig:one}. Clearly, the crossover occurs around $U_c/t\approx 11$, and the doping 
dependences of $\Delta_f$ and $\Delta_{SC}$ change dramatically, consistent with our discussion given above.
Fig. \ref{fig:two} presents the results with different $U/t$ as a function of $x$. It is remarkable to see a dome-like shape naturally appear in $\Delta_{SC}(x)$ for $U/t\geq 12$, and the 
'optimal doping' (the doping with largest $\Delta_{SC}$) is pushed to higher doping as $U$ increases, as expected.
We also plot out $Z$ and compare it with the Gutzwiller approximation $g_t=2x/(1+x)$. The slave spin method indeed matches the Gutzwiller approximation very well at small $x$, and 
the high order corrections due to the finite $U$ become significant at larger $x$.

We want to emphasize several advantages of the slave spin formalism compared to the traditional slave boson approach.
In the traditional slave boson approach, the charge degrees of freedom are represented by the slave bosons which are usually assumed to have Bose-Einstein condensate at mean-field level. 
As a result, the bosonic degrees of freedom are in fact treated classically. In the slave spin formalism, both the slave spin and the spinon are treated quantum mechanically at the mean-field
level already, which is the main reason why it can capture various exotic strongly correlated states at the mean-field level.
Second, although the slave boson approach can qualitatively obtain a quasiparticle weight $Z\sim x$, it is not straightforward to obtain the Gutzwiller factor $g_t$ directly from the 
Hubbard model in the large $U$ limit.
Recently, a new slave boson scheme that can correctly capture $g_t$ in the large $U$ limit has been proposed, but it has a serious drawback that the non-interacting limit can not be obtained.
\cite{lechermann2007}
In contrast, the mean-field equations based on the slave spin method are tailored to yield $g_t$ in the large $U$ limit, and the U(1) version can obtain the non-interacting limit correctly. 
As a result, it is crucial to use the slave spin formalism to study the crossover in the $t-J-U$ model discussed above, since it is necessary to treat both the strongly and the weakly 
interacting limits within the same framework.

{\it Pairing from charge interactions} --
We can use the same framework to study the superconductivity induced from the attractive nearest neighbor Coulomb interaction, $H_{C} = -V\sum_{<i,j>} n_i n_j$.
While $H_C$ can induce a $d$-wave superconducting state just like $H_J$, we find that the behavior of the superconducting state is fundamentally different.
Following the procedure of the slave spin formalism, $H_C$ can be rewritten as:
\be
H_{C} = -V\sum_{<i,j>} n_i n_j\approx -VZ^2\sum_{<i,j>} n^f_i n^f_j,
\label{hc}
\ee
where $n^f_i = \sum_\sigma f^\dagger_{i\sigma} f_{i\sigma}$.
If we introduce the same Hubbard-Stratonoch fields proposed by Ubbens and Lee\cite{ubbens1992} to decouple Eq. \ref{hc} in $d$-wave pairing channel, the mean-field pairing term is:
\be
H^{MF}_C = - \sum_{\vec{k}} Z^2\Delta^\prime(\vec{k}) f_{\vec{k}\uparrow} f_{-\vec{k}\downarrow} + h.c.,
\label{hcmf}
\ee
where $\Delta^\prime(\vec{k}) = V\Delta_f(\cos k_x - \cos k_y)$.
It can be seen that $H^{MF}_C$ has a renormalization of $Z^2$. 
Because the bandwidth of the spinon hopping term given in Eq. \ref{mf} is renormalized by $Z$ only, the pairing term is more suppressed  than 
the kinetic energy by a factor of $Z$, indicating that both $\Delta_f$ and $\Delta_{SC}$ would go to zero at small $x$ in large $U$.
As a result, the crossover observed in the case of the pairing driven by spin interactions will not happen in this case.

{\it Discussion} --
At the first sight, it seems that the crossover discussed above could be understood as a version of BEC-BCS crossover\cite{ranninger1996,chen1999,pieri2000,chen2005} 
due to the reduction of the bandwidth. However, to reach the BEC regime, 
the chemical potential $\mu$ has to be smaller than the minimum of the band energy so that no Fermi surface is left and the system becomes 'bosonic'. 
In our calculations, although the Fermi energy is reduced by $Z$, $\mu$ remains higher than the minimum of the band energy in every result, indicating that 
the fermionic pairing is still dominating. Physically, this can be understood as follows. The main effect of the Hubbard $U$ is to suppress charges from moving instead of providing extra glues 
for the pairing. Moreover, the Hubbard $U$ tends to eliminate local pairing since it costs a large energy to put two electrons on the same site.
Therefore, the pairing remains non-local despite of the kinetic energy being reduced by $U$, and the BEC picture does not work here.

To account for the phase transition at finite temperature in cuprates, an improvement beyond the mean-field level is necessary. In the present mean-field theory, 
$\Delta_f$ and $\Delta_{SC}$ have the same transition temperature as implied in Eq. \ref{dsc}. 
This is due to the assumption of the slave spin dynamics being the same for each site. 
Generally speaking, both the quantum and the thermal fluctuations could invalidate this assumption. 
As a result, $\langle S^-_{i\uparrow}S^-_{j\downarrow}\rangle$ could deviate from $Z$ beyond the mean-field level. 
If such a formalism beyond mean-field level is available, the transition temperatures of $\Delta_f$ and $\Delta_{SC}$ could be different. 
In this case, the superconducting phase is characterized by $Z<<1$, $\Delta_f\neq 0$, and $\langle S^-_{i\uparrow}S^-_{j\downarrow}\rangle \neq 0$, 
the pseudogap phase is characterized by $Z<<1$, $\Delta_f\neq 0$, and $\langle S^-_{i\uparrow}S^-_{j\downarrow}\rangle = 0$, and the strange metal phase is characterized by $Z<<1$, $\Delta_f= 0$, 
and $\langle S^-_{i\uparrow}S^-_{j\downarrow}\rangle = 0$. 
The present theory could be extended to the pairing state intertwinned with other orders like charge density wave 
state, stripe state, etc.

Another drawback of the present slave spin formalism is that the renormalization on the spinon is not directly considered through the mean-field equations. In addition to $g_t$, the Gutzwiller 
approximation also leads to a renormalization factor $g_s=4/(1+x)^2$ to the nearest neighbor antiferromagnetic Heisenberg interaction\cite{gutzwiller1963,florian1990}, and we 
have assumed $g_s=1$ in the present theory. 
Nevertheless, in the large $U$ limit, we can replace $J$ by $g_sJ$ in the $H_J$ term in the present slave spin formalism. 
Then it can be seen that $g_s J$ gets even bigger at small $x$, which simply makes the crossover more robust. 
As a result, we conclude that ignoring $g_s$ in the present theory would not undermine the conclusion of our results.
Ideally, one should obtain $H_J$ term via the superexchange mechanism with the inclusion of the slave spin and then derive the mean-field equations with 
both $H_J$ and $H_U$ terms treated equally. Improvement of the slave spin formalism to overcome problems mentioned above is still in progress.

{\it Conclusion}--
We have developed a BCS mean-field theory implemented with the slave-spin formalism and have investigated the superconducting states in the $t-J-U$ model. 
We have found that this formalism naturally predicts two distinct gaps, the spinon pairing gap $\Delta_f$ and the superconducting gap $\Delta_{SC} = Z\Delta_f$. 
For the case of the pairing arising from the spin interaction $H_J$, we have found that a crossover from strong to weak pairing states driven by the doping $x$ could occur as $U$ exceeds the 
critical value $U_c$ for the Mott insulating state at half-filling. In the strong pairing regime, $\Delta_f$ is largely enhanced due to the reduction of the spinon bandwidth, but 
the superconducting gap $\Delta_{SC}\sim x$ at small $x$. In the weak pairing regime, both $\Delta_f$ and $\Delta_{SC}$ behave similarly. 
On the other hand, if the pairing is induced by charge interactions, both $\Delta_f$ and $\Delta_{SC}$ would go to zero at small $x$, and consequently no crossover would occur. 
We have obtained a superconducting dome in ths phase diagram of $\Delta_{SC}$ vs $x$ with reasonable parameters, and all the results are obtained at the mean-field 
level. Our results have demonstrated that the slave spin formalism is a powerful technique to study study physics related to transition from strong to weak coupling states within the same framework.

{\it Acknowledgement}--
We appreciate fruitful discussions with T. K. Lee.
This work is supported by a start up fund from Binghamton University.

%\bibliography{slaves-sc}{}

\newpage
\section{Supplementary Materials}
Here we derive the mean-field equations from Eq. 4 in the main text. First, we derive the BCS gap equation in the spinon sector.
Using the Bogoliubov transformation, we obtain
\beq
f_{\vec{k}\uparrow}&=& \cos\frac{\theta_{\vec{k}}}{2} \alpha_{\vec{k},+} + \sin\frac{\theta_{\vec{k}}}{2} \alpha_{\vec{k},-},\nn\\
f^\dagger_{-\vec{k}\downarrow}&=& \sin\frac{\theta_{\vec{k}}}{2} \alpha_{\vec{k},+} - \cos\frac{\theta_{\vec{k}}}{2} \alpha_{\vec{k},-},\nn\\
\cos\theta_{\vec{k}}&=&\frac{\vert \langle\tilde{z}_{\sigma}\rangle\vert^2 \epsilon(\vec{k}) -\mu}{E(\vec{k})}\,\,\,,\,\,\,
\sin\theta_{\vec{k}}=\frac{\Delta(\vec{k})}{E(\vec{k})},\nn\\
E(\vec{k})&=&\sqrt{(\vert \langle\tilde{z}_{\sigma}\rangle\vert^2 \epsilon(\vec{k}) -\mu)^2+\Delta^2(\vec{k})}.
\eeq
$\alpha_{\vec{k},\pm}$ are the Bogoliubov quasiparticles with eigen energies $\pm E(\vec{k})$.
The self-consistent equations are:
\beq
\Delta_f&=& \frac{1}{2N}\sum_{\vec{k}} d(\vec{k})\sin\theta_{\vec{k}}\left[n_f(-E(\vec{k})) - n_f(E(\vec{k})) \right],\nn\\
n&=&1-x\nn\\
x&=& \frac{1}{N} \sum_{\vec{k}} \cos\theta_{\vec{k}}\left[n_f(-E(\vec{k})) - n_f(E(\vec{k})) \right],
\eeq
where $n_f(E)$ is the Fermi-Dirac function, and 
$x$ is the doping away from the half-filling.

The iteration procedure is as follows. Starting from an initial guess of $Z$, we first solve the BCS gap equation in the spinon sector subject to the fixed doping $x$. 
Then, we can compute the average kinetic energy of the spinon
\be
\epsilon^{\sigma\sigma} = \frac{1}{\Omega} \sum_{\vec{k}} \epsilon(\vec{k}) \cos\theta_{\vec{k}}\left[n_f(-E(\vec{k})) - n_f(E(\vec{k})) \right],
\ee
which will be used in the mean-field Hamiltonian for the slave spin $H^{s,MF}$.

The constraint to remove the enlarged Hilbert space is
\be
S^z_{i\sigma} = f^\dagger_{i\sigma}f_{i\sigma} - \frac{1}{2}.
\label{cons}
\ee
This constraint will be taken into account after the slave spin mean-field Hamiltonian is diagonalized.
Since the slave spin sector is effectively a single site problem in the mean-field theory, it can be diagonalized ecactly. 
Then we determine $\lambda$ by satisfying the constraint in Eq. \ref{cons} 'on the average' by
\be
M_\sigma\equiv \langle S^z_\sigma\rangle = \frac{1}{\Omega}\sum_{\vec{k}}\langle f^\dagger_{\vec{k}\sigma}f_{\vec{k}\sigma}\rangle - \frac{1}{2}
\ee
We can further simplify the above equation to
\be
1-x = M_\uparrow+M_\downarrow + 1\to M_\uparrow+M_\downarrow = -x.
\label{lambda}
\ee
Since we are only interested in the spin singlet pairing states, we have $M_\uparrow=M_\downarrow=-x/2$.

After $\lambda$ is determined, we can compute $Z$ which will be used for the next cycle of the calculation We repeat the procedure until the self-consistency is reached with an error less than 
$10^{-6}$.

\end{document}